# CONTRIBUTION OF INFORMATION AND COMMUNICATION TECHNOLOGY (ICT) IN COUNTRY'S H-INDEX

[1]MARYAM FARHADI[1*], HADI SALEHI[2,3], MOHAMED AMIN EMBI[2], MASOOD FOOLADI[1], HADI FARHADI[4], AREZOO AGHAEI CHADEGANI[1], and NADER ALE EBRAHIM[5]

[1]Department of Accounting, Mobarakeh Branch, Islamic Azad University, Mobarakeh, Isfahan, Iran

[2]Faculty of Education, Universiti Kebangsaan Malaysia (UKM), Bangi, 43600, Malaysia

[3]Faculty of Literature and Humanities, Najafabad Branch, Islamic Azad University, Najafabad, Isfahan, Iran

[4]School of Psychology and Human Development, Faculty of Social Sciences and Humanities, Universiti Kebangsaan Malaysia (UKM), Malaysia

[5]Department of Engineering Design and Manufacture, Faculty of Engineering, University of Malaya, Kuala Lumpur, Malaysia

[*]Corresponding author: Maryam Farhadi, Email: farhadim58@gmail.com)

## ABSTRACT

The aim of this study is to examine the effect of Information and Communication Technology (ICT) development on country's scientific ranking as measured by H-index. Moreover, this study applies ICT development sub-indices including ICT Use, ICT Access and ICT skill to find the distinct effect of these sub-indices on country's H-index. To this purpose, required data for the panel of 14 Middle East countries over the period 1995 to 2009 is collected. Findings of the current study show that ICT development increases the H-index of the sample countries. The results also indicate that ICT Use and ICT Skill sub-indices positively contribute to higher H-index but the effect of ICT access on country's H-index is not clear.

# 1. INTRODUCTION

In the last decades, technology has improved the life style and made easier the human activities through a new civilization. A new revolution named Information and Communication Technology (ICT) has occurred in the present world [1]. ICT is defined as a concept including information equipment as well as computers and software, auxiliary equipment that is connected to a computer (such as calculators, cash registers, photocopiers), communications equipment, and instruments [2]. In other words, ICT is a technology with the aim of connecting computers and other communication equipment to gather, produce, process, classify, manage, create, and distribute information. After the invention of machines in 1970s, industrial revolution has been started and machines have been used instead of handworks. However, ICT revolution and use of internet caused many structural changes [3]. As an example, the United States productivity has been revived during the late 1990s and the early 2000s because of the ICT development [4]. There are many studies introducing ICT as an important element to facilitate economic growth in both developing and developed countries [5, 6]. In addition, ICT has the potential to increase health systems through the designing new methods to prevent and detect disease [7]. Other studies confirm the dominant role of ICT in improving and modernizing the learning methods and educational systems [8, 9]. The fast progress in technology in the last two decades has reduced the cost of ICTs. In this situation, it is easier for more people to use and enjoy the advantages of ICTs [10].

Although, the key role of ICT is approved in different areas such as engineering [11], education [12], health care [13], and economy [6], its effect on the degree of scientific ranking of a country is yet unexplored. Therefore, the aim of this study is to explore the impact of ICT development as measured by ICT Development Index (IDI) on H-index as a proxy of scientific progress of a country.

In addition, in order to shed more lights on this area, present study applies three components of ICT development namely, ICT Skill, ICT Use and ICT Access to examine the distinct impact of these components on H-index as a proxy of scientific progress of a country. To this purpose, this study applies a multiple regression analysis on a panel of 14 Middle East countries over the period 1995 to 2009.

The current study suggests the use of H-index provided by SCImago database [14] to measure the scientific ranking of country. Based on Hirsch [15], a scientist or a group of scientists has H-index of h when h of their papers has at least h citations. In other words, H-index is the number of articles with citation number higher or equal to h [15]. The H-index, which has been invented by Hirsch in 2005, proposes a simple and apparently strong measure to compare the scientific productivity and visibility of countries and institutions [16].

H-index can measure the quantity and quality of papers simultaneously since the total number of published articles and the number of citations received by that articles are considered in the calculation of H-index. In addition, there are many studies, which reveal the benefits of H-index as an appropriate and fair tool for evaluating and ranking scientists, laboratories, scientific journals, universities, institutions and countries [17, 18, 19, 20, 21].

Hirsch [15] proposes that H-index can be used to appraise not only the scientists but also countries and institutions. Hence, following Jacso [22], current study also uses H-index as a proxy to measure scientific ranking of countries.

# 2. METHODOLOGY AND DATA

## 2.1. Empirical Model

In order to examine the effect of ICT development on scientific degree of 14 Middle East countries, this study applies the regression techniques with countries' H-index as dependent variable and IDI as independent variable as follows:

$$H_{it} = \alpha_0 + \alpha_1 IDI_{it} + \varepsilon_{it} \qquad (1)$$

where $\alpha$'s are parameters to be estimated, H represents H-index, IDI stands for ICT development index and subscripts $i$ and $t$ show the $i^{th}$ country in $t^{th}$ year, respectively. $\varepsilon$ is the error term which is assumed to be independent of IDI. Then, for more inspections, instead of IDI, which captures the composite effect of ICT development, the current study uses three ICT sub-indices including ICT Access, ICT Use and ICT Skill as explanatory variables. In this approach, we can test the distinct impact of each sub-index on the country's H-index. Consequently, in this step of the analysis, following equation (Equation 2) is used:

$$H_{it} = \mu_0 + \mu_1 Access_{it} + \mu_2 Use_{it} + \mu_3 Skill_{it} + \delta_{it} \qquad (2)$$

where $\mu$'s are parameters to be estimated, H shows H-index and Access, Use and Skill stand for ICT development sub-indices. Moreover, subscripts $i$ and $t$ indicate the $i^{th}$ country in $t^{th}$ year, respectively, and $\delta$ is the error term which is assumed to be independent of Access, Use and Skill variables.

**2.2. Data and Measurement**

**2.2.1. ICT development**

The IDI, as an indicator of ICT development is computed based on Principal Component Analysis (PCA), which combines various indicators in order to make single value. For the first time, IDI is introduced in 2009 by Measuring the Information Society [23] and considers the degree of progress in ICTs in more than 150 countries. Composition of this indicator includes ICT Access showing infrastructure and access, ICT Skill referring to ICT skill or capability and ICT Use displaying the extent of using ICT. Required ICT data to compute the IDI is extracted from United Nations Educational Scientific and Cultural Organization (UNESCO) and ITU. Nowadays, ICT is considered as an important factor in development of countries, which are moving towards knowledge or information-based societies. It is assumed that these countries have experienced the process of ICT development to turn into an information society. Therefore, the ITU [23] suggests the three-phase model including ICT readiness (showing the degree of networked infrastructure and ICT access), ICT intensity (showing the degree of ICT use by the society) and ICT impact (showing the result of effective and efficient ICT use). IDI is a composite indicator including ICT access, ICT skill and ICT use. Each sub-index is composed of a number of indicators explained as follow:

1. Access Sub-Index shows the availability of ICT infrastructure and individuals' access to basic ICTs. This sub-index has five indicators including fixed telephone lines per 100 inhabitants, mobile cellular telephone subscriptions per 100 inhabitants, international Internet bandwidth (bit/s) per Internet user, proportion of households with a computer and proportion of households with Internet access at home.

2. Use Sub-Index shows the real degree of ICTs strength of use based on the available infrastructure and access indicators. Considering the limitation on data for ICT Use at the global scale and results of the PCA, this subgroup has three indicators including Internet users per 100 inhabitants, fixed broadband Internet subscribers per 100 inhabitants and mobile broadband subscriptions per 100 inhabitants.

3. Skill Sub-Index captures the degree of ICT skill in different countries. Since the required data for many developing countries are not collected, an appropriate indicator can be the degree of literacy and education. Especially in developing countries, the poor level of education is a major obstacle to the effective use of internet and computers. With the inclusion of ICT in school courses, school attendance may offer an appropriate indicator for students' exposure to the Internet or computers. Therefore, this subgroup has three indicators including, adult literacy rate, secondary gross enrolment ratio and tertiary gross enrolment ratio. The required data for this subgroup are extracted from the UNESCO Institute for Statistics (UIS) [24].

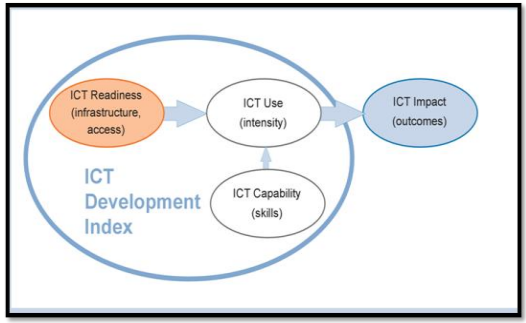

*Figure 1. Three stages in the evolution towards an information society (Source: ITU [25])*

### 2.2.2. h-index

Different sources report the H-index. Some of these sources are subscription-based sources such as Scopus and Web of Science (WoS) and some of them are free access sources such as Harzing's Publish or Perish program on the basis of Google Scholar entries. It should be noted that these sources report different H-index for the same institution, country and scholar because of their different coverage. For example, WoS covers a high range of published journals while it does not cover high impact conferences. Although, the coverage of publications in Scopus is poor prior to 1996, it covers conferences properly. Documents in Google Scholar receive more citations in comparison with those in Scopus and WoS. It is argued that Google Scholar has the most coverage of journals and conferences particularly those published after 1990 [26].

The current study collects the required data on H-index for the panel of 14 Middle East countries over the period 1995 to 2009, from SCImago database [14]. The SCImago Journal and Country Rank website presents the journal and country scientific indicators based on the information included in the Scopus database.

### 3. FINDING AND DISCUSSION

In this section, we first estimate Equation 1, which explores the impact of ICT development on H-index as a proxy of country's scientific ranking. Then, applying the three ICT sub-indices, this study presents the estimation results for Equation 2, which is corresponding to the effect of ICT development sub-indices on H-index.

*Table 1. Regression Results for the Impact of ICT development on H-index*

| Dependent Variable: H-index | | | |
|---|---|---|---|
| **Variables** | **Coefficient** | **Standard Error** | **T** |
| IDI | 31.69 | 4.67*** | 6.79 |
| Constant | 24.16 | 13.35* | 1.81 |
| Observation = 210 | | | |
| F(1, 208) = 46.07*** | | | |
| R-squared = 0.1813 | | | |
| Adjusted R-squared = 0.1774 | | | |
| Note: | | | |
| ***, ** and * denote statistically significant at 1%, 5% and 10%, respectively. | | | |
| Standard errors are heteroskedasticity consistent. | | | |

Table 1 shows the effect of ICT development on each country's H-index. The estimation results indicate a positive and significant effect of ICT development on country's scientific ranking as proxied by H-index. This finding suggests that suitable policies aim at improving the level of ICT development can significantly increase the scientific rankings of a country. The results also assert that if a country raises its ICT development index by one unit, the H-index will increase by 31 units. As can be seen in Table 1, the adjusted R-squared is equal to 0.1774, which means that ICT development can explain 17.74% of variations in country's H-index.

Table 2 exhibits the impact of ICT development sub-indices on H-index of 14 sample countries. The estimated coefficients show a positive and significant effect of ICT Use and ICT Skill on country's scientific ranking. This finding asserts that increasing the application of ICTs and improving the ICT skills might significantly increase the scientific degree of a country. The results also indicate that if a country increases the ICT Use and ICT Skill indices by one unit, its H-index will raise by 133 and 407 units, respectively. As can be seen in Table 2, the coefficient explaining the effect of ICT Access on H-index is negative but not significant. Finally, based on the adjusted R-squared, ICT sub indices can explain 27.11% of variations in country's H-index.

*Table 2. Regression Results for the Impact of ICT development sub-indices on H-index*

| Dependent Variable: H-index | | | |
|---|---|---|---|
| **Variables** | **Coefficient** | **Standard Error** | **t** |
| ICT Access | -25.46 | 49.80 | -0.51 |
| ICT Use | 133.23 | 64.64** | 2.06 |
| ICT Skill | 407.28 | 64.86*** | 6.28 |
| Constant | -151.10 | 35.82*** | -4.22 |
| Observation = 210 | | | |
| F(3, 206) = 26.91*** | | | |
| R-squared = 0.2816 | | | |
| Adjusted R-squared = 0.2711 | | | |
| Note: | | | |
| ***, ** and * denote statistically significant at 1%, 5% and 10%, respectively. | | | |
| Standard errors are heteroskedasticity consistent. | | | |

## 4. CONCLUSION AND IMPLICATIONS

This paper focuses on examining the effect of ICT development on the scientific degree of 14 Middle East countries as proxied by H-index. The results of the regression analysis show that ICT development has a positive and significant effect on the H-index of these countries. Moreover, applying the ICT development sub-indices, this study finds that ICT Use and ICT Skill might have positive effects on H-index but the effect of ICT Access is not clear since its corresponding coefficient is not statistically significant. Therefore, we can conclude that ICT induces outcomes that leads to higher H-index values and raises the scientific level of sample countries. In this situation, policy makers should aim increasing the level of ICT development through increasing its indicators including fixed and mobile telephone lines, international Internet bandwidth, proportion of households with a computer and Internet access, Internet users, fixed and mobile broadband Internet subscribers, adult literacy rate, secondary and tertiary gross enrolment ratios.

**AUTHOR PROFILES:**

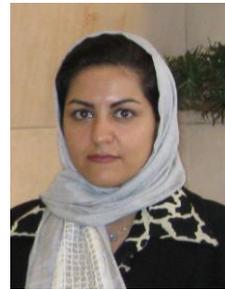

**Maryam Farhadi** received the Bachelor degree in pure mathematics from university of Isfahan in 2002 and master of Economics from Tarbiat Modarres University in 2006. Currently, she has finished her PhD in Economics in National University of Malaysia (UKM). Her interests are in Information and communication Technology, Economic growth and productivity.

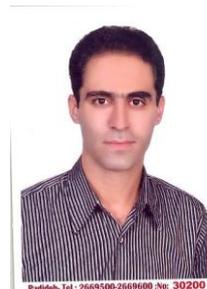

**Hadi Salehi** is a senior lecturer in the Faculty of Literature and Humanities, Najafabad Branch, Islamic Azad University, Iran where he teaches undergraduate and postgraduate courses in TESL. His main research interests include Language Learning Strategies, Materials Development, and Language Assessment, ICT, E-Learning, Mobile Learning, and Washback of High-stakes Tests.

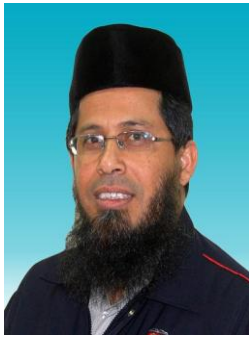

**Mohamed Amin Embi** is a Professor of Technology-enhanced Learning at the Faculty of Education, UKM. He is the recipient of the ISESCO Science Laureate 2010 & National Academic Award 2006.

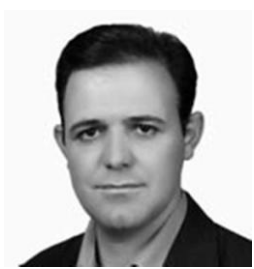

**Masood Fooladi** is a lecturer in Department of Accounting at Islamic Azad University, Mobarakeh Branch, Isfahan, Iran. He has graduated in master of Accounting from Tarbiat Modarres University, Tehran, Iran. Now, he is a PhD Candidate of Accounting at School of Accounting, Faculty of Economics and Management, National University of Malaysia (UKM), Selangor, MALAYSIA. He has around 7 years of teaching experience at university. He has around 10 research papers in various journals and conferences in financial accounting and corporate governance area.

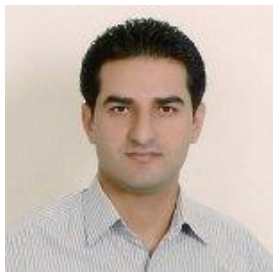

**Hadi Farhadi** is a Ph.D Candidate in Psychology School of Psychology and Human Development, Faculty of Social Sciences and Humanities National University of Malaysia. He is interested in industrial organizational psychology - consumer behavior psychology - personality traits - workplace deviant behavior and cyber deviant.

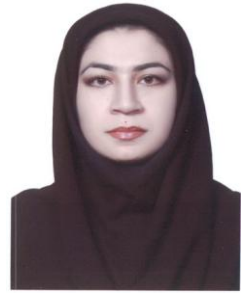

**Arezoo Aghaei Chadegani** received the Bachelor degree in accounting from Payam Noor university of Isfahan in 2003, and master of accounting from Islamic Azad University, Mobarakeh Branch in 2006. Currently, she is a PhD Candidate of accounting in National University of Malaysia (UKM). Her interests are in auditing, audit quality, financial accounting.

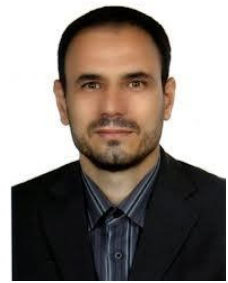

**Nader Ale Ebrahim** has Technology Management PhD degree from the Department of Engineering Design and Manufacture, Faculty of Engineering, University of Malaya. He holds a Master of Science in the mechanical engineering from University of Tehran with distinguished honors, as well as more than 17 years experience in the establishing R&D department in different companies, project director and project coordinator and Knowledge based system implemented in R&D department.

8